\documentclass{INTERSPEECH2023}

\interspeechcameraready

\newcommand{\bhline}{\noalign{\hrule height 1.5pt}} 

\usepackage{multirow}
\usepackage{arydshln}
\usepackage{here}

\title{Anomalous Sound Detection Based on Sound Separation}

\name{Kanta Shimonishi$^1$, Kota Dohi$^2$, Yohei Kawaguchi$^2$}
\address{
 $^1$Ritsumeikan University, Japan \leavevmode \\
 $^2$Hitachi, Ltd., Japan 
}
\email{is0460rf@ed.ritsumei.ac.jp,\{kota.dohi.gr, yohei.kawaguchi.xk\}@hitachi.com}

\begin{document}

\maketitle
 
\begin{abstract}
This paper proposes an unsupervised anomalous sound detection method using sound separation. 
In factory environments, background noise and non-objective sounds obscure desired machine sounds, making it challenging to detect anomalous sounds. 
Therefore, using sounds not mixed with background noise or non-purpose sounds in the detection system is desirable.
We compared two versions of our proposed method, one using sound separation as a pre-processing step and the other using separation-based outlier exposure that uses the error between two separated sounds. 
Based on the assumption that differences in separation performance between normal and anomalous sounds affect detection results, a sound separation model specific to a particular product type was used in both versions.
Experimental results indicate that the proposed method improved anomalous sound detection performance for all Machine IDs, achieving a maximum improvement of 39\%.
\end{abstract}

\section{Introduction}
Anomalous sound detection (ASD) is a technique for identifying whether an observed sound is normal or anomalous~\cite{Dohi_arXiv2022_02}. 
This technique makes it possible to detect anomalous operating sounds when a machine malfunctions and helps in monitoring the machine's condition. 
It is challenging to collect anomalous-sound data in the real world because such sounds are rare~\cite{Koizumi_DCASE2020_01}. 
Anomalous sounds also have a wide range of sound variations, and there is likely to be a large amount of unknown data, making supervised learning of ASD difficult. 
Therefore, ASD is usually conducted using an unsupervised method that uses only normal sounds during training. 
During inference, the anomaly degree is calculated on the basis of how well the observed sounds fit the learned distribution. 
If the anomaly exceeds a pre-defined threshold, it is judged as an anomalous sound.
In the Detection and Classification of Acoustic Scenes and Events (DCASE) 2020 Challenge Task 2~\cite{Koizumi_DCASE2020_01}, several ASD methods using outlier exposure (OE) were proposed. 
In OE, neural networks classify the Machine-type and Machine-ID of input sounds~\cite{Vinayavekhin2020,Zhou2020,Giri2020,Lopez2020}, detect the machines' motion segments~\cite{nishida2022anomalous}, and calculate the anomaly degree in accordance with their accuracy. 
ASD with OE is based on the assumption that unknown anomalous sounds that have not been used for training are challenging for classifying the machine type (Machine-type) or product type (Machine-ID) or to detect operating intervals.\par

In a factory environment, machines other than the target machine are in operation~\cite{kawaguchi2019anomaly}. 
The observed sounds include noise, which decreases the performance of ASD. 
If the difference between normal and anomalous sounds is slight, ASD becomes even more difficult. 
Therefore, it is considered adequate to remove noise and non-target sounds from the observed sounds and use them for ASD. 
A semi-supervised non-negative matrix factorization (NMF)~\cite{lee2000algorithms} method of extracting the target machine sound using a pre-trained basis of environmental noise is used to pre-process ASD~\cite{aiba2021noise}. 
However, machine sound is difficult to define independence, sparsity, and low rankness, which may make sound separation (SS) difficult when using methods such as NMF. 
Poor separation accuracy can have a significant impact on later ASD results.
In addition, when performing ASD, a human must listen to and analyze the machine sounds.
Therefore, separating the target machine sounds is necessary.\par
We propose an ASD method that uses deep learning for SS. 
There are two versions of this method. 
The first version uses deep-learning-based SS as a pre-processing step for conventional unsupervised ASD and detects anomalous sounds from the separated sounds. 
Since only normal sounds are used when training the SS model, the SS model can separate normal sounds. 
If the model is trained to separate for a specific Machine-ID, even for the same Machine-type, separation becomes problematic if the target Machine-ID is different, and the separation of anomalous sounds, which are an unknown domain, becomes more difficult. 
It is predicted that normal sounds will be cleanly denoised. 
In contrast, anomalous sounds will not be sufficiently denoised or distorted, and the different separation between normal and anomalous sounds is expected to be used for ASD. 
The second version uses OE based on SS. 
In OE based on SS, Two SS models are used: one for a specific Machine-type and the other for a specific Machine-ID. 
The anomaly degree is calculated on the basis of the error of the separated sounds output from each model.

\begin{figure*}[h]
\centering
\centering
\scalebox{1.03}{
 \begin{minipage}{.5\textwidth}
 \centering
 \scalebox{0.03}{
 \includegraphics[]{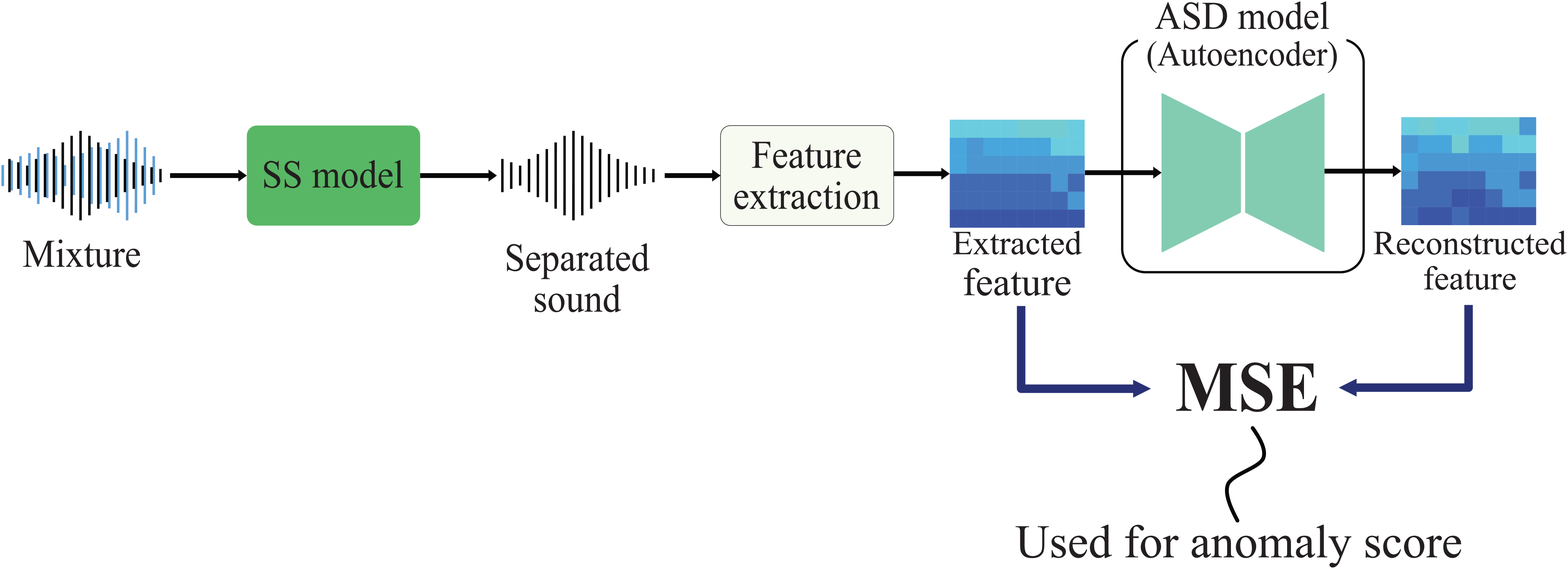}
 }
 \caption{Overview of ASD after SS}
 \label{fig:ASD_twostep}
 \end{minipage}
 \begin{minipage}{.5\textwidth}
 \centering
 \scalebox{0.03}{
 \includegraphics[]{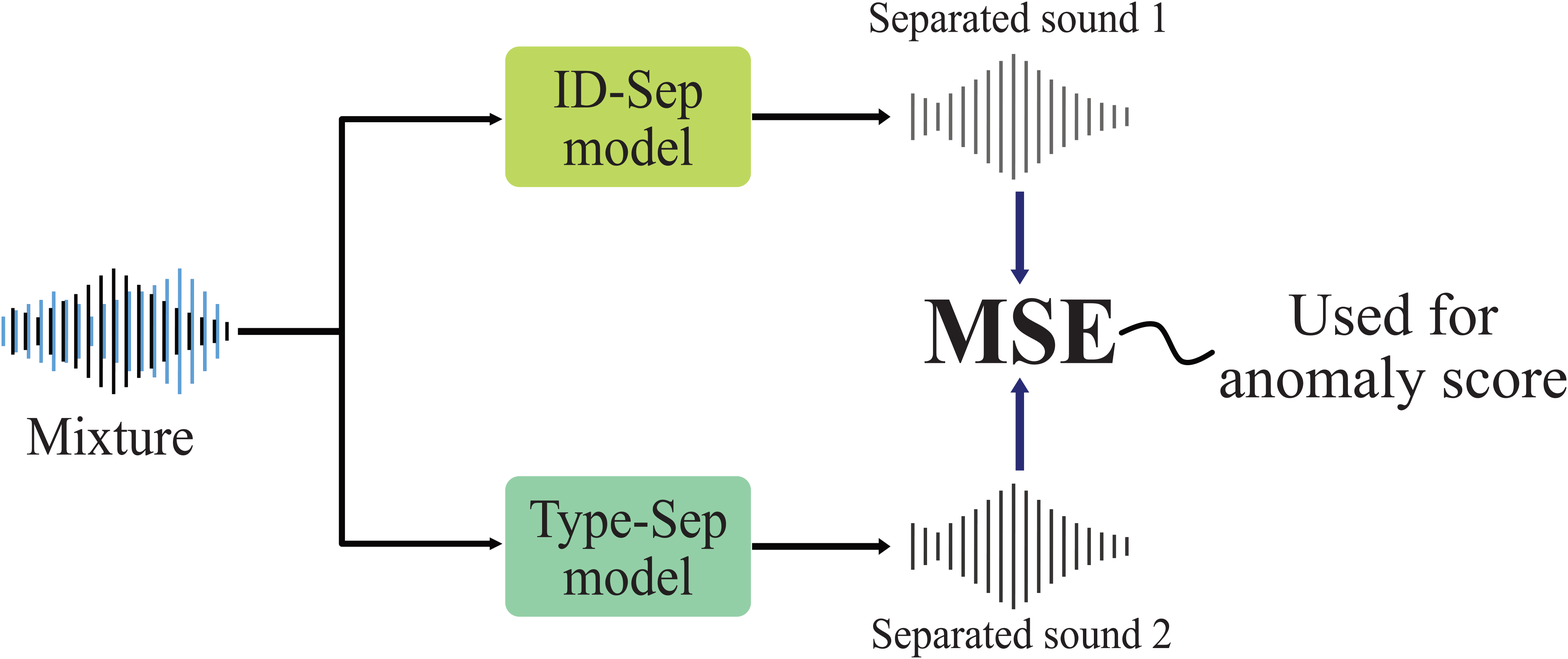}
 }
 \vspace{13pt}
 \caption{ASD using separation-based OE}
 \label{fig:ASD_OurlierExposure}
 \end{minipage}
 }
\end{figure*}

\section{Problem description}
\subsection{Unsupervised anomalous sound detection}
ASD calculates the anomaly level on the basis of the acoustic feature $\textbf{X}$ of the observed sound. 
The anomaly calculator $\mathcal{A}$ with parameter $\theta$ outputs the anomaly degree $\mathcal{A}_{\theta}(\textbf{X})$. 
If $\mathcal{A}_{\theta}(\textbf{X})$ is larger than the preset threshold $\phi$, it is judged as an anomaly with the following equation.
\begin{equation}
\label{ASD}
\mathrm{Decision}=\left \{ \begin{array}{ll}
 \mathrm{Anomaly} & (\mathcal{A}_{\theta}(\textbf{X}) > \phi) \\
 \mathrm{Normal} & (\mathrm{otherwise})
\end{array} \right.
\end{equation}
Unsupervised ASD is based on outlier detection, and methods using autoencoders (AEs)~\cite{koizumi2019batch} have been widely used. 
AEs learn encoders and decoders so that the difference between the input and output of the neural network (reconstruction error) is reduced.
When a normal sound is input, the reconstruction error becomes small, but when an anomalous sound is input, the reconstruction error becomes large.
The input and output errors are used to calculate the anomaly degree and execute ASD.

\subsection{Sound separation}
SS is a technique for separating a sound mixture into its individual sources~\cite{belouchrani1997blind}. 
Most of the research on SS has focused on speech~\cite{wang2018supervised,zhang21v_interspeech} and music~\cite{defossez2019music,hennequin2020spleeter}. 
Universal sound separation (USS)~\cite{kavalerov2019universal,tzinis2020improving,munakata2021multiple} has been attracting attention because it targets various sounds, not limited to speech and music. 
Masked-based methods are mainly used in USS. 
In masked-based SS, a mask-prediction network estimates the mask by encoding the $\boldsymbol{x}\in\mathbb{R}^L$ mixture with signal length $L$ and inputting it to the mask-prediction network. 
The mask process separates the encoded mixtures, and the decoder resynthesizes the separated source $\boldsymbol{y}_{n}\in\mathbb{R}^L$~\cite{koizumi2021df}. 
Let $n=1,2,\cdots,N$ denote the index of the sound source and $N$ the number of sound sources. 
The separation procedure is formulated as
\begin{equation}
\label{masked_SS}
\boldsymbol{y}_{n} = \mathrm {Dec}(\mathrm {Enc}(\boldsymbol{x}) \odot \mathcal M_{n}(\mathrm {Enc}(\boldsymbol{x}))),
\end{equation}
where $\mathrm {Enc}(\cdot),\mathrm {Dec}(\cdot)$ are the encoder and decoder, respectively, $\odot$ is the element-wise multiplication, and $\mathcal M_{n}(\cdot)$ is a mask predicted by the network for each source. 
Various networks have been proposed as masked-based SS models~\cite{kavalerov2019universal,luo2019conv,tzinis2020sudo}.

\section{Proposed method}
Previous studies have stated that it is difficult to obtain ground truth in advance~\cite{aiba2021noise}. 
This study assumes that it is possible to record ground truth in situations where there is little or no background noise, such as before the start of factory operations.
This assumption is made because recording sound with only the target machine is possible since no other machines operate before the plant starts.

\subsection{Anomalous sound detection after sound separation}
With this version of the proposed method, two models, SS and ASD are serially connected to calculate the anomaly level directly from the input sound. 
By isolating the target machine sound before ASD, clean sound without noise can be used for ASD, and it may be possible to detect slight differences between normal and anomalous sounds.
A schematic of this version is shown in Figure \ref{fig:ASD_twostep}. 
The mixed sound is the first input, and the machine sound separated using the SS model in the previous stage is the output. 
The acoustic features extracted from the separated sounds are input to the ASD model in the second stage to determine whether the sound is normal or anomalous.\par
The SS and ASD models are trained independently and connected only during inference. 
The training of the SS model requires data on the combination of the mixture and ground truth of the target machine sound. 
The model takes a mixture of sounds and trains the output-separated sounds so that they approach the ground truth. 
SS is used as a pre-processing step for ASD, so the ASD model is trained using the sound mixtures as with the conventional method that does not use SS. 
The ASD model learns so that the output reconstructed from the sound mix is close to the input. In both SS and ASD models, only normal sounds are used during training.

\begin{figure*}[h]
\centering
\scalebox{0.2}{
 \includegraphics[]{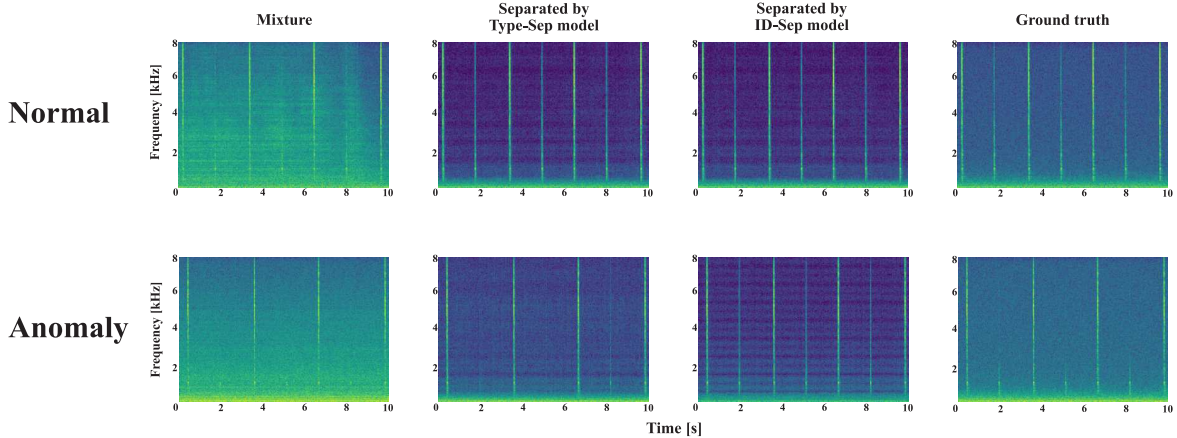}
 }
 \caption{Example spectrograms for section 02 of valve. }
 \label{fig:Spectrogram}
\end{figure*}
\subsection{Separation-based outlier exposure}
The separation-based OE version of the proposed method uses two SS models: Machine-ID-wise Separation (ID-Sep) model, which separates normal sounds of a specific Machine-ID, and Machine-type-wise Separation (Type-Sep) model, which separates normal sounds of a specific Machine-type. 
An overview of this version is shown in Figure \ref{fig:ASD_OurlierExposure}. 
The mixture is input to the ID-Sep and Type-Sep models, and the separated sound is output from each model.
On the basis of the error between the two separated sounds, the anomaly degree of the input sound is calculated. 
The neural network only executes SS and uses the observed SS results. 
It is possible to separate normal sounds in the learned known domain, but separating anomalous sounds in the unknown domain is challenging. 
It is expected that the results of the separation of normal and anomalous sounds can be used for ASD.
\par
The ID-Sep model overfits the normal sound of a particular Machine-ID, so the performance of separating anomalous sounds should significantly degrade. The Type-Sep model overfits a particular Machine-type regardless of the Machine-ID, so the separation performance of the anomalous sound is lower. However, it should be lower than with the ID-Sep model. There is no significant difference between the models in separation performance for normal sounds, which are also used for training. In other words, if the error of each separation is small, the sound is judged to be normal, and if the error is significant, it is judged to be anomalous. The error between the two separated signals is calculated using the mean squared error (MSE).\par
The ID-Sep and Type-Sep models are trained independently and on different data since they have different separation goals. The ID-Sep model targets a specific Machine-ID, so it is trained to remove the noise and sounds of other Machine-IDs. Therefore, it is trained on data that contain a mixture of Machine-IDs that are not the target Machine-IDs. Since the Type-Sep model targets a specific Machine-type regardless of the Machine-ID, it only needs to remove noise. To train the Type-Sep model, data in which only noise is mixed are used. The dataset is described in Section 4.1.
\begin{table}[h]
\centering
\caption{Structure of autoencoder}
\label{table:AE}
\vspace{-4pt}
\scalebox{1.0}{
\begin{tabular}{l|l}
\bhline
\multirow{5}{*}{Encoder} & Linear (in:640 , out:128) \\
 & Linear (in:128 , out:128) \\
 & Linear (in:128 , out:128) \\
 & Linear (in:128 , out:128) \\
 & Linear (in:128 , out:8) \\ \hline
\multirow{5}{*}{Decoder} & Linear (in:8 , out:128) \\
 & Linear (in:128 , out:128) \\
 & Linear (in:128 , out:128) \\
 & Linear (in:128 , out:128) \\
 & Linear (in:128 , out:640) \\ \bhline
\end{tabular}
}
\end{table}

\section{Experiment}
\subsection{Dataset}
Although the mixture sound and corresponding ground truth are necessary for learning and evaluating SS, the currently available datasets for anomaly detection do not include ground truth. 
Therefore, we created a dataset that contains clean machine sounds and factory environmental noise recorded during the creation of the dataset MIMII DG~\cite{Dohi2022} for ASD.
The target machine sounds were slider and valve, and the Machine-ID data of sections 00, 01, and 02 of these components were used for each Machine-type. 
The SNR of the mixing was randomized from \{-5, 0, 5\}dB.\par

The data for training the ID-Sep model consisted of three mixed patterns. For example, when section 00 was the target Machine-ID, the following was used, and the same for the other Machine-IDs.
\begin{enumerate}
\setlength{\leftskip}{0.6cm}
   \item section 00 + noise
   \item section 00 + section 01 + noise
   \item section 00 + section 02 + noise
\end{enumerate}
There were 990 data items for each pattern, creating a total of 2,970 data items for one Machine-ID.\par
The Type-Sep model training data mixes only noise with the target machine sound, as in the first mixing pattern of the ID-Sep model training data.
There were 990 data items for each Machine-ID, and 2,970 data items were created for one Machine-type. \par

The evaluation data were the same as the Type-Sep training data in terms of the mixing pattern. A total of 100 data items were created, 50 normal sounds and 50 anomalous sounds, for each Machine-ID.

\subsection{Training and evaluation setup}
\begin{table*}[h]
\centering
\caption{SI-SDRi for each separation model [dB]}
\label{table:separation}
\vspace{0pt}
\scalebox{1.0}{
  \begin{minipage}{.48\textwidth}
  \captionsetup{labelformat=empty,labelsep=none,font=footnotesize}
  \caption{(a) Slider}
  \vspace{-4pt}
    \centering
    \scalebox{0.95}{
      \begin{tabular}{l||cc:cc:cc}
    \bhline
    & \multicolumn{2}{c:}{\textbf{section 00}} & \multicolumn{2}{c:}{\textbf{section 01}} & \multicolumn{2}{c}{\textbf{section 02}} \\
    & N & A & N & A & N & A  \\ \hline
    Type Sep & 5.15 & 2.99 & 5.35 & 3.70 & 4.60 & -0.48 \\
    ID Sep & 4.84 & -0.46 & 4.70 & -0.36 & 6.37 & 0.95 \\
    \bhline
  \end{tabular}
      }
  \end{minipage}
  \begin{minipage}{.48\textwidth}
  \captionsetup{labelformat=empty,labelsep=none,font=footnotesize}
  \caption{(b) Valve}
  \vspace{-4pt}
    \centering
    \scalebox{0.95}{
      \begin{tabular}{l||cc:cc:cc}
    \bhline
    & \multicolumn{2}{c:}{\textbf{section 00}} & \multicolumn{2}{c:}{\textbf{section 01}} & \multicolumn{2}{c}{\textbf{section 02}} \\
    & N & A & N & A & N & A \\ \hline
    Type Sep & 12.30 & 7.48 & 6.50 & 6.32 & 10.20 & 6.22 \\
    ID Sep & 11.64 & 6.22 & 6.58 & 5.79 & 10.37 & -0.32 \\
    \bhline
  \end{tabular}
      }
  \end{minipage}
  }
\end{table*}
\begin{table*}[h]
\setcounter{table}{2}
\centering
\caption{AUC with SS [\%]}
\vspace{-7pt}
\label{table:ASDwithSS}
\centering
\scalebox{1.0}{
 \begin{minipage}{.47\textwidth}
 \captionsetup{labelformat=empty,labelsep=none,font=footnotesize}
 \caption{(a) Slider}
 \vspace{-4pt}
 \centering
 \scalebox{0.95}{
 \begin{tabular}{l||ccc|c}
 \bhline
 & \multicolumn{3}{c|}{Machine-ID} & \multirow{2}{*}{\textbf{Avg}} \\
 & \multicolumn{1}{c}{\textbf{00}} & \multicolumn{1}{c}{\textbf{01}} & \textbf{02} & \\
 \hline
 Baseline & 50.36 & 55.52 & 58.60 & 54.83 \\
 \hdashline
 After-Type-Sep & 57.08 & 77.32 & 20.72 & 51.71 \\
 After-ID-Sep & 77.08 & \textbf{92.84} & \textbf{89.56} & \textbf{86.49} \\
 Separation-based OE & \textbf{85.08} & 79.92 & 73.32 & 79.44 \\
 \bhline
 \end{tabular}
 }
 \end{minipage}
 \begin{minipage}{.47\textwidth}
 \captionsetup{labelformat=empty,labelsep=none,font=footnotesize}
 \caption{(b) Valve}
 \vspace{-4pt}
 \centering
 \scalebox{0.95}{
 \begin{tabular}{l||ccc|c}
 \bhline
 & \multicolumn{3}{c|}{Machine-ID} & \multirow{2}{*}{\textbf{Avg}} \\
 & \multicolumn{1}{c}{\textbf{00}} & \multicolumn{1}{c}{\textbf{01}} & \textbf{02} & \\
 \hline
 Baseline & 50.72 & 55.60 & 56.60 & 54.31 \\
 \hdashline
 After-Type-Sep & 78.00 & 58.60 & 67.56 & 68.05 \\
 After-ID-Sep & \textbf{79.44} & 56.96 & \textbf{96.00} & \textbf{77.47} \\
 Separation-based OE & 14.48 & \textbf{59.24} & 91.64 & 55.12 \\
 \bhline
 \end{tabular}
 }
 \end{minipage}
 }
\end{table*}

We use Conv-TasNet~\cite{luo2019conv} as the SS model. Conv-TasNet is an end-to-end SS model that uses sound waveforms for input and output. The number of output sources is set to 1, and the other model structures are based on the best-performing version of \cite{luo2019conv}. The batch size is 2, Adam optimizer is used, and the learning rate is 0.0001. The loss function uses the L1 loss of the output source and ground truth, and the model was trained for 50 epochs per Machine-ID for the ID-Sep model and per Machine-type for the Type-Sep model. Separation performance was evaluated on the basis of scale-invariant signal-to-distortion ratio improvement (SI-SDRi), which is the difference between the SI-SDR of the estimated sound and ground truth and that of the mixture and ground truth. The SI-SDR is calculated as
\begin{equation}
\label{si-sdr}
\textrm{SI-SDR}(\boldsymbol{s},\hat{\boldsymbol{s}})=10\log_{10} \frac{\lVert \alpha \boldsymbol{s} \rVert^2}{\lVert \alpha \boldsymbol{s} - \hat{\boldsymbol{s}} \rVert^2},
\end{equation}
where $\boldsymbol{s}$ is the ground truth and $\hat{\boldsymbol{s}}$ is the estimated signal or mixture. Also, $\alpha=\langle \boldsymbol{s},\hat{\boldsymbol{s}}\rangle/\lVert\boldsymbol{s}\rVert^2$ and $\langle \boldsymbol{s},\hat{\boldsymbol{s}}\rangle$ are the inner products of $\boldsymbol{s}$ and $\hat{\boldsymbol{s}}$, respectively.\par

The ASD model uses AEs and trained on the same data as the Type-Sep model. The sound waveforms are transformed into a log-Mel spectrogram with a frame size of 1024, hop size of 512, and 128 Mel bins. The five frames are concatenated to generate a 640-dimensional feature vector that is input to the autoencoder. The structure of the autoencoder is shown in Table \ref{table:AE}. Batch normalization and a rectified linear unit are inserted after each linear layer except the final layer. The batch size was 512, Adam optimizer was used, and the learning rate was 0.001. We used the MSE loss of the input and output vectors for the loss function and trained 100 epochs for each Machine-type. ASD performance was evaluated on the basis of the area under the receiver operating characteristic curve (AUC), which is defined as
\begin{equation}
\label{AUC}
\textrm{AUC}=\frac{1}{N_{n}N_{a}}\sum_{i = 1}^{N_{n}}\sum_{j = 1}^{N_{a}}\mathcal{H}(\mathcal{A}_{\theta} (\boldsymbol{x}_j^a)-\mathcal{A}_{\theta} (\boldsymbol{x}_i^n)),
\end{equation}
where $\mathcal{H}(x)$ returns 1 if $x > 0$, and 0 otherwise. Here, $\{\boldsymbol{x}_i^n\}_{i = 1}^{N_{n}}$ and $\{\boldsymbol{x}_j^a\}_{j = 1}^{N_{a}}$ are the test data for normal and anomalous sounds, respectively, and $N_{n}$ and $N_{a}$ are the number of normal and anomalous-sound test data items, respectively.

\subsection{Experimental results}
The SS results are listed in Table \ref{table:separation}. 
In Table \ref{table:separation}, N and A mean normal and anomalous sounds, respectively.
Both the slider and valve performed well in separating normal sounds for many Machine-IDs. Since only normal sounds are used in training, even sounds of the same Machine-type are challenging to separate if they are anomalous sounds. 
When section 02 of the valve was separated using the ID-Sep model, there was a 10.69dB difference in separation performance between normal and anomalous sounds. 
Comparing the separation performance between the ID-Sep and Type-Sep models, there was not much difference in the performance for normal sound, but there was a significant difference for anomalous sound. 
However, in section 01 of the valve, the difference in separation performance between normal and anomalous sounds and between the ID-Sep and Type-Sep models was slight. 
The reasons may be that learning is complex or that the difference between normal and anomalous sounds is slight. Example spectrogram for section 02 of the valve are shown in Figure \ref{fig:Spectrogram}.
\par
The results of ASD with SS are shown in Table \ref{table:ASDwithSS}. "After-Type-Sep" and "After-ID-Sep" signify the version of the proposed method that uses SS as a pre-processing step for ASD, using the Type-Sep and ID-Sep models, respectively. 
Baseline was a conventional method that does not use SS. 
In most cases, the AUCs of the proposed versions were better than that of the baselines. 
Both proposed versions improved ASD performance for sections 00 and 01 of the slider and sections 01 and 02 of the valve, and section 02 of the valve improved ASD performance by 39.4\% at most. 
These results indicate that using separated machine sounds improves ASD performance. However, for section 02 of the slider, the ASD performance decreased with After-Type-Sep. 
Table \ref{table:separation} shows that the SS performance of section 02 of the slider is relatively low for normal sounds with Type-Sep. 
Therefore, it is considered that distortion occurs in the separated normal sounds, and the anomaly degree calculated during ASD increases, resulting in false detection as an anomalous sound. For section 00 of the valve, the separation-based OE version reduced ASD performance. 
Table \ref{table:separation} shows a significant difference in the separation performance of normal sounds between ID-Sep and Type-Sep for the corresponding Machine-ID, and a slight difference between the two SS models with high separation performance for anomalous sounds. 
From the above, it can be seen that if the difference in separation performance between normal and anomalous sounds is not as expected, the proposed ASD method will be negatively affected.
Considering the similarity between normal and anomalous sounds, it is necessary to train the separation model so that anomalous sounds, which are unknown domains, can hardly be separated.
Comparing only the versions of the proposed method, the score of After-ID-Sep was the highest in most cases.
The separation performance of anomalous sounds was deficient compared with that of normal sounds because separation was executed using a model specialized for a particular Machine-ID. 
The distortion caused by separating anomalous sounds may improve ASD performance because the anomaly degree is increased. 
In addition, either After-ID-Sep or Separation-based OE achieved the highest score for every Machine-ID.
Thus, the two versions of the proposed method are complementary.

\section{Conclusion}
We proposed an ASD method using SS. 
We compared two versions of the proposed method, one that uses deep-learning-based SS as a pre-processing step for ASD and the other that uses separation-based OE, as well as a baseline that does not use SS. 
Experimental results indicate that the proposed method performed better than the baseline for many Machine-IDs. 
In particular, the version with After-ID-Sep improved ASD performance for all Machine-IDs.
We also found that the two versions are complementary, as the maximum score was achieved in After-ID-Sep or separation-based OE for all Machine-IDs.

\bibliographystyle{IEEEtran}
\bibliography{mybib}

\end{document}